\documentstyle[12pt,aaspp4,flushrt]{article}
\begin{document}
\parindent=1.0cm

\title{The Evolved Red Stellar Contents of the Sculptor Group 
Galaxies NGC55, NGC300, and NGC7793}

\author{T. J. Davidge \altaffilmark{1}}

\affil{Canadian Gemini Project Office, Herzberg Institute of 
Astrophysics, \\ National Research Council of Canada, 5071 W. Saanich Road, \\
Victoria, BC, Canada V8X 4M6\\ {\it email:tim.davidge@hia.nrc.ca}}

\and

\affil{Department of Physics and Astronomy, University of British Columbia, \\
Vancouver, BC Canada V6T 1Z1}

\altaffiltext{1}{Visiting Astronomer, Cerro Tololo Inter-American Observatory. 
CTIO is operated by AURA Inc., under contract to the NSF.}

\begin{abstract}

	Deep $J, H,$ and $K$ images are used to probe the evolved stellar 
contents in the central regions of the Sculptor group galaxies NGC55, NGC300, 
and NGC7793. The brightest stars are massive red supergiants (RSGs) with $K 
\sim 15 - 15.5$. The peak RSG brightness is constant to 
within $\sim 0.5$ mag in $K$, suggesting that NGC55, NGC300, and NGC7793 are at 
comparable distances. Comparisons with bright RSGs in the Magellanic Clouds 
indicate that the difference in distance modulus with respect to the LMC is 
$\Delta \mu = 7.5$. A rich population of asymptotic giant branch (AGB) stars, 
which isochrones indicate have ages between 0.1 and 10 Gyr, 
dominates the $(K, J-K)$ color-magnitude diagram (CMD) of each galaxy. 
The detection of significant numbers of AGB stars with ages near 10 
Gyr indicates that the disks of these galaxies contain an underlying old 
population. The CMDs and luminosity functions reveal significant 
galaxy-to-galaxy variations in stellar content. Star-forming activity in the 
central arcmin of NGC300 has been suppressed for the past Gyr with respect to 
disk fields at larger radii. Nevertheless, comparisons 
between fields within each galaxy indicate that star-forming activity during 
intermediate epochs was coherent on spatial scales of a 
kpc or more. A large cluster of stars, which isochrones suggest has 
an age near 100 Myr, is seen in one of the NGC55 fields. 
The luminosity function of the brightest stars in this cluster is 
flat, as expected if a linear luminosity-core mass relation is present. 

\end{abstract}

\section{INTRODUCTION}

	Recent studies of the structural characteristics of spiral galaxies 
suggest that the disk and spheroid do not evolve in isolation, but interact 
throughout the lifetime of a galaxy (e.g. Andredakis, 
Peletier, \& Balcells 1995). The observational signatures of 
these interactions are readily apparent in late-type 
spiral galaxies. For example, Courteau, de Jong, \& Broeils 
(1996) find a scale-free Hubble sequence among late-type spirals, a 
result which could be explained if the central light concentrations 
\footnote{ The presence of a traditional bulge in late-type spirals galaxies 
has been challenged by Bothun (1992) and Regan \& Vogt (1994), who concluded 
that the `bulge' in M33 is actually the central extension 
of the halo. However, Minniti, Olszewski, \& Rieke (1993) and Mighell \& Rich 
(1995) resolved the innermost regions of M33 into stars, and detected a 
significant intermediate age population. Minniti {\it et al.} (1993) argue 
that, on the basis of stellar content alone, M33 contains a 
central component that is distinct from the halo. Given this debate, in the 
current paper the term `central light concentration' is used to refer 
to what has traditionally been called the `bulge' in late-type spirals.} 
formed after disks. WFPC1 images discussed by Phillips {\it et al.} (1996) 
reveal that the central light concentrations of late-type spirals are 
structurally distinct from the bulges of early-type spirals, 
suggesting differences in evolutionary histories. 

	Surveys of the bright stellar content in nearby galaxies provide a 
direct means of studying the evolution of their central regions.
Photometric studies of first ascent and asymptotic giant branch 
(AGB) populations are of particular interest, as these stars 
probe evolution during early and intermediate epochs, when the basic 
properties of the disk and spheroid were being imprinted. While efforts 
to resolve the central light concentrations of these systems into stars require 
near-diffraction limited image quality to overcome crowding, the inner disks 
of many nearby systems can easily be resolved into stars from the ground. 
Although the Local Group contains the closest, most obvious objects for 
stellar content surveys, the number of targets is limited to three 
morphologically diverse spiral galaxies: the Milky-Way, M31, and M33. 
This limited sample makes it necessary to study more distant 
systems and, as the nearest collection of galaxies outside the Local Group, 
the Sculptor Group offers a number of lucrative targets that can be 
resolved from the ground.

	In the current paper, deep $J, H$, and $K$ images are used to 
investigate the photometric properties of cool stars in the inner disks of the 
Sculptor galaxies NGC55, NGC300, and NGC7793. The morphological types and 
integrated brightnesses of these galaxies, as assigned by Sandage \& Tammann 
(1987), are summarized in Table 1. Throughout this study it is assumed that 
these galaxies are equidistant with $\mu_0 = 26.0$, as derived for 
NGC300 by van den Bergh (1992) from a number of different standard candles. 
The Galactic reddening towards these galaxies is negligible (Burstein \& 
Heiles 1984). 

	There are a number of advantages to conducting photometric surveys of 
luminous, cool evolved stars at wavelengths longward of $1\mu$m. Not only is 
the contrast between bright cool stars and fainter unresolved objects in the 
disk enhanced at infrared wavelengths, but it is also possible to overcome the 
effects of line blanketing, which can affect the spectral energy 
distributions of moderately metal-rich giants at optical wavelengths 
(e.g. Bica, Barbuy, \& Ortolani 1991), and complicate efforts to derive 
bolometric corrections. In addition, near-infrared two-color 
diagrams can also be used to identify foreground stars, contamination from 
which may be significant at faint visible magnitudes. 

	NGC55 and NGC300 have been the targets of earlier photometric 
investigations. Deep broad- and narrow-band surveys of NGC55 (Pritchet {\it et 
al.} 1987) and NGC300 (Richer, Pritchet, \& Crabtree 1985; Zijlstra, Minniti, 
\& Brewer 1996) have revealed that the outer disks of these galaxies contain 
rich AGB populations. A comparison of the AGB luminosity functions suggests 
that the star-forming histories of NGC55 and NGC300 
during intermediate epochs were similar, but not 
identical (Pritchet {\it et al.} 1987). Freedman (1984) and Pierre \& Azzopardi 
(1988) used $B$ and $V$ photometry to survey the bright young stellar 
content of NGC300, while Kiszhurno-Kozrey (1988) used CCD observations to 
construct $(V, B-V)$ CMDs of two fields in NGC55. There is no published 
photometric study of the stellar content of NGC7793, although Catanzarite 
{\it et al.} (1995) report the discovery of Cepheids in this galaxy. The 
only published infrared survey of these galaxies was carried out by 
Humphreys \& Graham (1986), who obtained $JHK$ aperture measurements of 
red supergiant (RSG) candidates in NGC300. Spectroscopy revealed that almost 
half of the candidate objects were cool Galactic main sequence stars. 

	The paper is structured as follows. The observations, reduction 
techniques, and methods used to measure stellar brightnesses are discussed in 
\S 2. The luminosity functions (LFs), two-color diagrams (TCDs), 
and color-magnitude diagrams (CMDs) derived 
from these data are presented and compared in \S 3. In \S 4 the data 
are used to search for radial population gradients in NGC300 and NGC7793. 
A summary of the results follows in \S 5.

\section{OBSERVATIONS, REDUCTIONS, AND PHOTOMETRIC MEASUREMENTS}

	The data were recorded at the Cassegrain focus of the CTIO 1.5 metre 
telescope during the nights of UT July 19 -- 23 1996 
using CIRIM -- the facility near-infrared camera. 
Each pixel on the $256 \times 256$ Hg:Cd:Te array subtended 
0.6 arcsec on a side, so that the total imaged field was $2.6 \times 2.6$ 
arcmin. A complete observing sequence consisted of three co-added 60 sec 
integrations per filter recorded at four dither positions, defining a $5 
\times 5$ arcsec square on the sky. This sequence was repeated up to 
five times for each field to improve the signal-to-noise ratio. While the 
central regions of the galaxies were the primary targets for this program, 
fields at intermediate radii were also observed in NGC55 and NGC300. 
Additional details of the observations, including the locations of field 
centers, are listed in Table 2.

	The initial stages of the data reduction consisted of: 
(1) dark subtraction, (2) division by dome 
flats, and (3) subtraction of the DC sky levels, which were 
computed on a frame-by-frame basis using the DAOPHOT (Stetson 1987) 
sky routine. Additional calibration frames showing thermal artifacts
in $H$ and $K$ were constructed from the sky-subtracted data 
by stacking the images in these filters 
and then computing the median signal at each 
pixel location. The result was subtracted from each of the 
sky-subtracted $H$ and $K$ images. The processed images for each field 
were then aligned and the median at each pixel location computed. 
The final median-combined $K$ images, trimmed to the area common to the 
four dither positions, are shown in Figures 1 -- 5.

	Stellar brightnesses were measured with the PSF-fitting routine 
ALLSTAR (Stetson \& Harris 1988), which is part of the DAOPHOT photometry 
package. Aperture corrections were derived from PSF stars after neighboring 
objects were subtracted from the images. Unresolved stars in 
each galaxy introduce non-uniformities in the background, which complicate 
efforts to obtain reliable photometry. The iterative technique described 
by Davidge, Le F\`{e}vre, \& Clark (1991) was used to model and remove this 
component.

	The photometric calibration was defined with 25 observations of 
standard stars from the lists compiled by Elias {\it et al.} (1982) and 
Casali \& Hawarden (1992). Given that extinction coefficients in the 
infrared are small and require a large number of standard star observations 
covering a range of airmasses, it was decided to fix these 
coefficients at the values derived by Guarnieri, Dixon, \& Longmore (1991) for 
Mauna Kea. The method of least squares was then used 
to fit linear transformation equations of the form:

\vspace{0.3cm}
\hspace*{4.0cm} M = $\mu \times C + m + \zeta$
\vspace{0.3cm}

\noindent{where} M and $m$ are the standard and instrumental magnitudes, and 
$C$ is the instrumental color, to the extinction-corrected measurements.

\section{THE PHOTOMETRIC PROPERTIES OF RESOLVED STARS}

\subsection{$JHK$ Luminosity Functions and Data Completeness}

	The $J, H,$ and $K$ LFs for all five fields are compared in Figure 6. 
The decrease in counts at the faint end indicates that incompleteness becomes 
significant when $J = 19.5 - 20.0$, $H = 19.0 - 19.5$, and $K = 18 - 18.5$. 
Sample completeness depends on crowding, which can change 
significantly over small angular scales in the central regions of galaxies; 
consequently, these completeness limits should be considered to be 
representative only.

	When examined on a field-by-field basis, the $J, H,$ and $K$ LFs show 
very good agreement. Moreover, with the possible exception of 
NGC55 Field 2, the logarithmic LFs follow linear trends at the bright end, 
so they can be characterised by a power-law exponent. Power-law exponents were 
derived by performing least squares fits at the bright end of each LF, 
and the results derived from the $K$ LFs in the interval $K = 16 - 18$ are 
listed in Table 3. The exponents vary significantly from galaxy-to-galaxy, 
indicating differences in stellar content. Furthermore, the power-law 
exponents derived for the two fields in NGC55 and NGC300 agree within 
the estimated uncertainties, suggesting that the bright red stellar 
content within these galaxies does not change markedly with radius.

\subsection{The $(K, J-K)$ CMDs and $(J-H, H-K)$ TCDs}

	The $(K, J-K)$ CMDs of the five fields are shown in Figure 7. 
The CMDs contain a mixture of RSGs, AGB stars, and 
foreground objects, and in this section an effort is made to identify each 
of these components.

	Each galaxy contains a population of stars 
with $K \sim 15 - 15.5$ and $J-K \sim 1$, and these objects 
are likely bright RSGs. Evidence to support this interpretation comes from the 
infrared aperture measurements of the optically brightest red stars in NGC300 
obtained by Humphreys \& Graham (1986). It is apparent from Figure 7 that 
the objects in their sample of spectroscopically 
confirmed RSGs have brightnesses and colors that are 
in excellent agreement with the brightest stars in the 
various Sculptor fields. The peak brightness of the RSG clump in each galaxy 
occurs near $K \sim 15 - 15.5$, indicating that NGC55, NGC300, and NGC7793 have 
roughly comparable distances. NGC300 Field 1 is of interest as there are 
no stars with $K \leq 16$, indicating that 
recent star formation has not occured in this field.

	When $K \geq 16$ both RSGs and AGB stars are present, and 
infrared observations of similar stars in Local Group galaxies can be used 
to define the photometric characteristics of these objects. 
The CMDs of RSGs and long period variable (LPV) AGB stars in 
the LMC and SMC, taken from the photometric compilations of Elias, Frogel, \& 
Humphreys (1985) and Wood, Bessell, \& Fox (1983), are shown in Figure 8. The 
brightest RSGs in the Magellanic Clouds have M$_K \sim -11.5$, 
while the Magellanic Cloud LPV sequence peaks near $M_K \sim -10$. There is 
also a clear color separation between RSGs and AGB LPVs near $J-K \sim 1$.

	Also shown in Figure 8 is the composite CMD of all 5 Sculptor fields, 
assuming a common distance modulus of $\mu_0 = 26$. There 
are many similarities between the Sculptor and Magellanic Cloud CMDs. 
For example, the brightest stars in Sculptor and the Magellanic Clouds have 
comparable M$_K$ and $(J-K)_0$, indicating that the RSGs with $K \sim 
15 - 15.5$ in the Sculptor fields are likely near the peak RSG brightness. 
In fact, this agreement supports a difference in distance modulus $\Delta 
\mu \sim 7.5$ between the LMC and the three Sculptor galaxies. 
Furthermore, the redward extent of the AGB sequence in Sculptor grows towards 
fainter brightnesses, and a similar trend is seen in the Magellanic Cloud 
data. Finally, the Sculptor CMD shows a discontinuity near M$_K \sim -10$, 
which corresponds to the AGB-tip brightness in the Magellanic Clouds. The 
comparison in Figure 8 thus indicates that the stellar contents of the Sculptor 
galaxies and the Magellanic Clouds are similar.

	The $(J-H, H-K)$ TCD provides a powerful means of 
assessing foreground contamination, and can also be used to isolate 
RSG and AGB stars. The $(J-H, H-K)$ TCDs for the various fields are shown 
in Figure 9. The sequences defined by Galactic dwarfs with spectral-types 
G -- M (Bessell \& Brett 1988), LMC and SMC M supergiants (Elias, Frogel, \& 
Humphreys 1985), and LMC LPVs (Wood {\it et al.} 
1983; Wood, Bessell, \& Paltoglou 1985) are also plotted on Figure 9. 
The RSGs studied by Humphreys \& Graham (1986) fall along the LMC/SMC 
supergiant sequence on the $(J-H, H-K)$ TCD. 

	Foreground dwarfs have $J-H \leq 0.7$, and it is apparent that the 
majority of objects detected in the Sculptor fields have near-infrared spectral 
energy distributions (SEDs) that are very different from solar 
neighborhood dwarfs. This is significant since Humphreys \& Graham (1986) 
found that four of the nine bright RSG candidates identified from optical 
data were actually foreground dwarfs; the comparisons in Figure 9 suggests that 
foreground contamination is much less significant in bright star samples 
selected at infrared wavelengths.

	The TCDs confirm that all five fields contain a significant population 
of objects with colors appropriate for RSGs. However, there is also a 
population of stars in each field with $J-H$ colors that are too large to be 
RSGs, and the locus of these objects parallels the LMC LPV sequence.

	Bertelli {\it et al.} (1994) modelled the evolution of stars from 
the main sequence to the AGB-tip for a broad range of masses and chemical 
compositions. Solar metallicity AGB sequences with log(t) = 8.0, 9.0, and 10.0, 
constructed by connecting the RGB and AGB-tip data summarized in Table 11 of 
Bertelli {\it et al.} (1994), are compared with the observed CMDs in Figures 7 
and 8.  $J-K$ colors were computed from the published $V-K$ values using the 
relation for solar neighborhood giants given by Frogel \& Whitford (1987).

	The comparisons between the isochrones and Magellanic Cloud data in 
Figure 8 are of particular interest, as the stars are spectroscopically 
confirmed members of the Magellanic Clouds, so that foreground contamination is 
not an issue, while the photometric measurements have little noise. The blue 
envelope and peak brightness of LPVs in the Magellanic Cloud are both 
matched by the log(t) = 8.0 sequence. However, the agreement between the models 
and the SMC observations is much poorer when M$_K \geq -8$, 
as there is a large number of stars redder than the log(t) 
= 10.0 sequence. Nevertheless, while there is difficulty matching the colors of 
the reddest objects, all but two of the stars with $(J-K)_0 
\geq 1.5$ have M$_K$ below the predicted log(t) = 10 AGB-tip brightness.
Comparisons with models from Bertelli {\it et al.} (1994) 
having other metallicities give similar results. 

	The isochrones indicate that the disks of NGC55, NGC300, and NGC7793 
contain stars that span a range of ages, as would be expected if the star 
formation rate has been roughly constant during intermediate epochs (Kennicutt, 
Tamblyn, \& Congdon 1994). NGC300 is unique, as the majority of stars in this 
field have ages in excess of 1 Gyr.

	The oldest stars are of interest for probing the age 
and early evolution of the disk. Sommer-Larson (1996) modelled the star-forming 
and chemical enrichment histories of disks, and concluded that 
the disks of early and late-type spiral galaxies have 
different ages, in the sense that late-type systems are younger. 
Although relatively faint, AGB-tip stars with log(t) = 10 still fall 
above the 100\% completeness level in most fields, 
and a significant population of AGB stars with 
ages approaching 10 Gyr are present in all Sculptor fields. 

	The NGC55 Field 2 CMD contains a well-defined 
sequence with $J-K \sim 1$ and $K \leq 17.5$. 
The stars that fall along this sequence are located near the 
eastern edge of Figure 2, and appear to belong to a richly populated, 
moderately young star cluster. The 100 Myr AGB isochrone runs parallel 
to, and to the red of, this sequence on Figure 7, and peaks $\sim 0.4$ 
mag in $K$ fainter than the observations, suggesting that the 
cluster may be slightly younger than 0.1 Gyr.

	The LF of the brightest stars in this relatively young, richly 
populated cluster is of interest for probing the advanced stages of stellar 
evolution. Early efforts to model evolution on the AGB by Paczynski (1970) 
revealed a linear relation between surface luminosity and core mass, which in 
turn implies a flat LF (Renzini 1977). Subsequent studies of AGB evolution 
employing improved input physics have confirmed the linear nature of the 
luminosity -- core mass relation for stars near the AGB-tip 
(e.g. Figure 7 of Boothroyd \& Sackmann 1988).

	The sequence in NGC55 Field 2 is scattered about a single $J-K$ 
color, so the stars should have, at least to first order, similar bolometric 
corrections; hence, the $K$ LF should be morphologically similar to 
the bolometric LF. The $K$ LF of stars in NGC55 Field 2 with $K \leq 17.5$ 
and $0.7 \leq J-K \leq 1.2$ is shown in the top panel of Figure 10; the portion 
of the CMD showing the AGB sequence and the color selection boundaries is 
shown in the lower panel. The solid line in the upper panel is the mean 
value of $n$ derived from all 9 LF points, and it is apparent that 
the individual datapoints can be matched by a constant. Consequently, 
the LF is consistent with a linear relation between luminosity and core mass, 
as predicted by stellar evolution theory. 

\subsection{The Bolometric LF}

	The bolometric LF provides an additional means of 
comparing the star-forming histories of galaxies. 
Bolometric LFs derived from infrared data are of particular interest for 
studies of very cool stars as the optical photometric properties 
of these objects are affected by line blanketing, which complicates 
efforts to compute accurate bolometric corrections. 

	$K$-band bolometric corrections, BC$_K$, were computed 
for stars in the $(K, J-K)$ CMDs using the relation between BC$_K$ and 
$J-K$ for field giants given in Figure 1b of Frogel \& Whitford (1987). The 
plotted relation terminates at $(J-K)_0 \sim 1.2$, and was 
extended to redder colors by paralleling the bulge giant relation 
and applying a 0.1 magnitude offset, based on the difference between the two 
relations at bluer $J-K$ colors. It should be noted that the bolometric 
corrections computed in this manner apply only to giants, so there 
is some uncertainty in the luminosities derived for supergiants.

	The bolometric LFs are compared in Figure 11, and 
galaxy-to-galaxy differences in stellar content are clearly evident. For 
example, the NGC300 LFs have a discontinuity at M$_{bol} \leq -6$, which is not 
seen in NGC55 and NGC7793, while the deficiency of luminous stars in 
NGC300 Field 1, noted in the previous section, is also apparent. 
Kolmogoroff-Smirnoff tests confirm that the NGC55 and NGC300 LFs differ 
in excess of the 95\% confidence level when M$_{bol} \leq -5$, which is the 
appromximate completeness limit of these data. The NGC55 and NGC7793 LFs are 
not significantly different.

	The LMC and SMC data plotted in Figure 8 indicate that color 
information can be used to isolate AGB LPVs in each galaxy. The bolometric LFs 
of stars with $(J-K) \geq 1.1$ are plotted as dashed lines in Figure 11 and, 
with the exception of NGC55 Field 1, there is excellent agreement between the 
LFs of the various fields when M$_{bol} \leq -5$. This similarity, 
which was verified with Kolmogoroff-Smirnoff tests, suggests that 
the star-forming histories of the three galaxies have been similar
when averaged over moderately long time scales. Richer {\it et al.} (1984) and 
Pritchet {\it et al.} (1987) used $V$ and $I$ data to derive AGB LFs for 
NGC300 and NGC55 and, with the exception of a difference in the number of 
objects with M$_{bol} \sim -6$, also found them to be very similar. 

\section{RADIAL POPULATION GRADIENTS IN NGC300 AND NGC7793}

	Galaxy formation models suggest that metallicity gradients form 
early-on in disks (Steinmetz \& Muller 1994, 1995), and studies of HII 
regions indicate that these trends are sustained during subsequent episodes 
of star formation (e.g. Zaritsky, Kennicutt, \& Huchra 1994 and references 
therein). It is not clear what effect galaxy morphology has on 
disk population gradients, although radial variations in the ratio of 
bulge-to-disk stars (Arimoto \& Jablonka 1991) will introduce 
apparent gradients in stellar content. Nevertheless, studies of HII regions in 
galaxies spanning a range of brightnesses and morphologies suggests that 
local properties, such as surface brightness, rather than global properties, 
such as morphology, dominate disk evolution (Ryder 1995). Such a dependence 
may explain why, for a given Hubble type, the size and direction of disk color 
gradients show considerable scatter (e.g. Terndrup et al. 1994).

	NGC300 and NGC7793 have orientations that are well suited for 
investigating the radial behaviour of disk stellar content; however, this is 
not the case for the edge-on system NGC55, so this galaxy is not considered 
in this Section. Population gradients may occur 
over relatively small spatial intervals close to the centers of galaxies and, 
in an effort to determine if population gradients are evident among the 
brightest red stars in NGC300 and NGC7793, the stellar contents in two 
annuli, centered on the inner spheroid of each galaxy, were compared. 
The annuli considered in NGC300 span the radial intervals $0 - 43$ (Ring 1) and 
$43 - 61$ (Ring 2) arcsec, while in NGC7793 the corresponding intervals are 
$0 - 36$ (Ring 1) and $36 - 55$ (Ring 2) arcsec. These intervals were 
selected so that each annulus contained comparable numbers of stars.

	The CMDs for Rings 1 and 2 in each galaxy are compared in Figure 
12, while the corresponding $K$ LFs are compared in the top two rows of 
Figure 13. The Ring 2 measurements in the right hand 
panel of Figure 13 have been shifted along the vertical axis to compensate 
for differences in surface brightness, based on the 
surface brightness profiles measured by Carignan (1985). These 
surface-brightness corrected data will serve as the basis for the 
comparisons discussed below.

	It is evident from Figure 12 that the stars in NGC300 Ring 2 may be 
slightly redder on average than those in NGC300 Ring 1. In fact, when 
$K \leq 18$ (ie. the approximate 100\% completeness level) $\overline{J-K} 
= 1.21 \pm 0.04$ in Ring 1, compared with $1.37 \pm 0.06$ in Ring 2. 
These values differ at roughly the $2-\sigma$ level, so the color difference is 
only marginally significant. For comparison, the mean $J-K$ 
colors for Ring 1 and 2 in NGC7793 are in excellent agreement. 

	The Ring 1 and 2 LFs in both galaxies do not differ significantly 
when $K \leq 18$. Although significant differences appear to occur 
when $K \geq 18$, these are likely due to the radial variation 
in the completeness level. When considered together, the CMDs and LFs 
in Figures 12 and 13 suggest that the bright stellar contents of NGC300 and 
NGC7793 do not change markedly with radius within an arcmin of the galaxy 
centers.

	NGC300 Fields 1 and 2 are separated by 3 arcmin, and the LFs 
of these areas are compared in the lower panel of Figure 13. As noted 
previously, Field 2 contains an excess population of bright stars with respect 
to Field 1, although the two LFs are in good agreement when $K \geq 16.5$. It 
appears that there has been considerable spatial coherence in the star-forming 
properties of NGC300 over scales of a few arcmin, which 
corresponds to $\sim 1$ kpc, during intermediate 
epochs. However, during the past Gyr star formation 
in the inner regions of NGC300 has been suppressed with respect to the 
outer disk.

\section{SUMMARY}

	Moderately deep near-infrared images have been used to probe the 
red stellar contents of the Sculptor group galaxies NGC55, NGC300, and NGC7793. 
The $(J-H, H-K)$ diagram constructed from these data demonstrates that 
surveys of the most evolved objects conducted in the infrared are 
less prone to foreground star contamination than their visible counterparts. 
All 3 galaxies contain a population of bright 
RSGs with $K \sim 15 - 15.5$, indicating that NGC55, NGC300, and NGC7793 
have comparable distances. Comparisons with RSGs in the Magellanic Clouds 
suggest that the difference in distance modulus between these galaxies and 
the LMC is $\Delta \mu \sim 7.5$. All 5 fields contain rich 
intermediate-age populations spanning a range of ages from 1.0 to at least 10 
Gyr. Therefore, the disks of these galaxies all contain 
an old stellar substrate. When averaged over timescales in excess of 1 Gyr or 
more, the star-forming histories of fields within a given galaxy show spatial 
coherence over kpc scales. However, this has not been the case during more 
recent epochs. For example, the central arcmin of NGC300 contains only a modest 
population of stars with ages less than 1 Gyr, indicating that recent star 
formation has been suppressed with respect to disk fields at larger radii. 
Moreover, despite having similar morphological characteristics, the 
star-forming histories of NGC55, NGC300, and NGC7793 during recent epochs show 
differences that are clearly apparent when comparing CMDs and LFs, 
although these differences become less pronounced when considering 
intermediate-age populations. NGC55 Field 2 contains a richly population 
cluster with an age $\sim 0.1$ Gyr. The LF of this cluster is flat, and 
hence is consistent with a linear relation between core mass and luminosity.

\pagebreak[4]
\begin{center}
TABLE 1

GALAXY PROPERTIES
\end{center}

\begin{center}
\begin{tabular}{lcr}
\hline\hline
NGC & Type & M$_B$ \\
\hline
55 & Sc & --20.1 \\
300 & ScII & --18.6 \\
7793 & Sd(s)IV & --18.8 \\
\hline
\end{tabular}
\end{center}

\pagebreak[4]
\begin{center}
TABLE 2

DETAILS OF OBSERVATIONS
\end{center}

\begin{center}
\begin{tabular}{lccccc}
\hline\hline
NGC & Field & RA & Dec & Exposure Time & FWHM \\
 & \# & (1950) & (1950) & (sec) & (arcsec) \\
\hline
55 & 1 & 00:12:24.1 & --39:28:03 & 3600 & 1.8 $(J)$, 1.5 $(HK)$ \\
 & 2 & 00:30:49.1 & --39:29:57 & 1440 & 1.5 $(JHK)$ \\
 & & & & & \\
300 & 1 & 00:52:30.8 & --37:57:26 & 1440 & 1.8 $(JHK)$ \\
 & 2 & 00:52:30.5 & --37:54:26 & 720 & 1.5 $(JHK)$ \\
 & & & & & \\
7793 & 1 & 23:57:49.2 & --32:35:26 & 4080 & 1.5 $(JK)$, 1.8 $(H)$ \\
\hline
\end{tabular}
\end{center}

\pagebreak[4]
\begin{center}
TABLE 3

POWER-LAW EXPONENTS FOR $K$ LFs
\end{center}

\begin{center}
\begin{tabular}{lcc}
\hline\hline
NGC & Field & $x$ \\
 & \# & \\
\hline
55 & 1 & $0.49 \pm 0.05$ \\
 & 2 & $0.48 \pm 0.06$ \\
 & & \\
300 & 1 & $1.02 \pm 0.07$ \\
 & 2 & $0.85 \pm 0.11$ \\
 & & \\
7793 & 1 & $0.78 \pm 0.06$ \\
\hline
\end{tabular}
\end{center}

\pagebreak[4]
\parindent=0.0cm
\begin{center}
REFERENCES
\end{center}

Andredakis, Y. C., Peletier, R. F., \& Balcells, M. 1995, MNRAS, 275, 874

Arimoto, N., \& Jablonka, P. 1991, A\&A, 249, 374

Bertelli, G., Bressan, A., Chiosi, C., Fagotto, F., \& Nasi, E. 1994, A\&AS, 
\linebreak[4]\hspace*{1.0cm}106, 275

Bessell, M. S., \& Brett, J. M. 1988, PASP, 100, 1134

Bica, E., Barbuy, B., \& Ortolani, S. 1991, ApJ, 382, L15

Bothun, G. D. 1992, AJ, 103, 104

Boothroyd, A. I., \& Sackmann, I.-J. 1988, ApJ, 328, 641

Burstein, D., \& Heiles, C. 1984, ApJS, 54, 33

Carignan, C. 1985, ApJS, 58, 107

Casali, M., \& Hawarden, T. 1992, JCMT-UKIRT Newsletter, 4, 33

Catanzarite, J. H., Horowitz, I. K., Freedman, W. L., \& Madore, B. F. 1995, 
\linebreak[4]\hspace*{1.0cm}BAAS, 27, 1294

Courteau, S., de Jong, R. S., \& Broeils, A. H. 1996, ApJ, 457, L73

Davidge, T. J., Le F\`{e}vre, O., \& Clark, C. C. 1991, ApJ, 370, 559

Elias, J. H., Frogel, J. A., \& Humphreys, R. M. 1985, ApJS, 57, 91

Elias, J. H., Frogel, J. A., Matthews, K., \& Neugebauer, G. 1982, AJ, 
\linebreak[4]\hspace*{1.0cm}87, 1029

Freedman, W. L. 1984, ApJ, 299, 74

Frogel, J. A., \& Whitford, A. E. 1987, ApJ, 320, 199

Guarnieri, M. D., Dixon, R. I., \& Longmore, A. J. 1991, PASP, 103, 675

Humphreys, R. M., \& Graham, J. A. 1986, AJ, 91, 522

Kennicutt, R. C. Jr., Tamblyn, P., \& Congdon, C. W. 1994, ApJ, 435, 22
 
Kiszkurno-Koziej, E. 1988, A\&A, 196, 26

Lacey, C. G., \& Fall, S. M. 1985, ApJ, 290, 154

Mighell, K. J., \& Rich, R. M. 1995, AJ, 110, 1649

Minniti, D., Olszewski, E. W., \& Rieke, M. 1993, ApJ, 410, L79

Paczynski, B. 1970, Acta Astron., 20, 47

Pierre, M., \& Azzopardi, M. 1988, A\&A, 189, 27

Phillips, A. C., Illingworth, G. D., MacKenty, J. W., \& Franx, M. 1996, 
\linebreak[4]\hspace*{1.0cm}AJ, 111, 1566

Pritchet, C. J., Richer, H. B., Schade, D., Crabtree, D., \& Yee, H. K. C., 
\linebreak[4]\hspace*{1.0cm}1987, ApJ, 323, 79

Regan, M. W., \& Vogel, S. N. 1994, ApJ, 434, 536

Renzini, A. 1977, in Advanced Stages in Stellar Evolution, ed. P. Bouvier, 
\linebreak[4]\hspace*{1.0cm}\& A. Maeder, (Geneva Observatory: Geneva), pp. 149

Richer, H. B., Pritchet, C. J., \& Crabtree, D. R. 1985, ApJ, 298, 240

Rieke, G. H., \& Lebofsky, M. J. 1985, ApJ, 288, 618

Ryder, S. D. 1995, ApJ, 444, 610

Sandage, A., \& Tammann, G. A. 1987, A Revised Shapley-Ames Catalog of 
\linebreak[4]\hspace*{1.0cm}Bright Galaxies, Carnegie, Washington.

Sommer-Larson, J. 1996, ApJ, 457, 118

Steinmetz, M., \& Muller, E. 1994, A \& A, 281, L97

Steinmetz, M., \& Muller, E. 1995, MNRAS, 276, 549

Stetson, P. B. 1987, PASP, 99, 191

Stetson, P. B., \& Harris, W. E. 1988, AJ, 96, 909

Terndrup, D. M., Davies, R. L., Frogel, J. A., DePoy, D. L., \& Wells, L. A. 
\linebreak[4]\hspace*{1.0cm}1994, ApJ, 432, 518

van den Bergh, S. 1992, PASP, 104, 861

Wood, P. R., Bessell, M. S., \& Fox, M. W. 1983, ApJ, 272, 99

Wood, P. R., Bessell, M. S., \& Paltoglou, G. 1985, ApJ, 290, 477

Zaritsky, D., Kennicutt Jr., R. C., \& Huchra, J. P. 1994, ApJ, 420, 87
 
Zijlstra, A. A., Minniti, D., Brewer, J. 1996, Messenger, 85, 23

\pagebreak[4]

\begin{figure}
\plotone{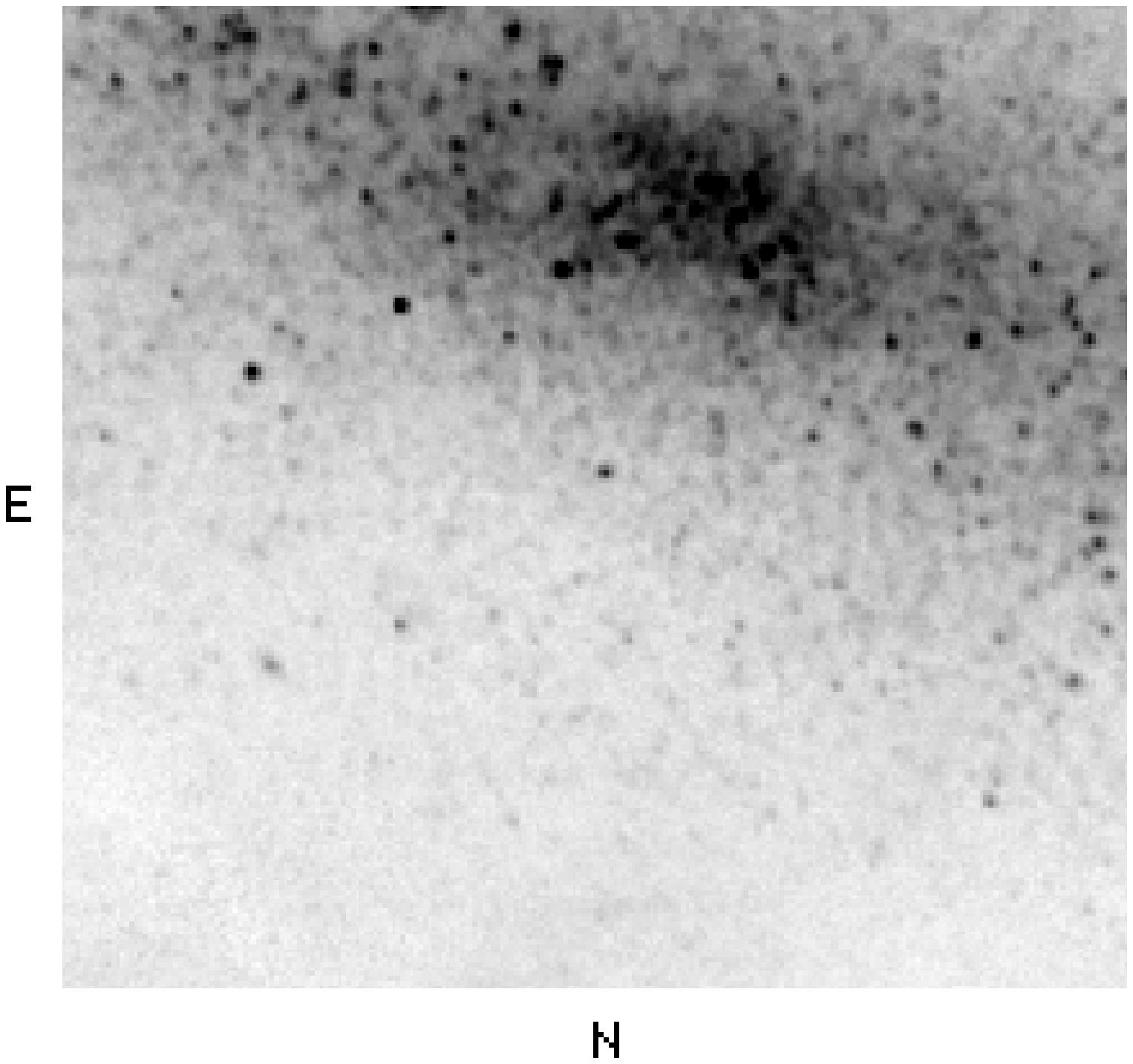}
\caption{Final $K$ image of NGC55 Field 1.}
\end{figure}

\begin{figure}
\plotone{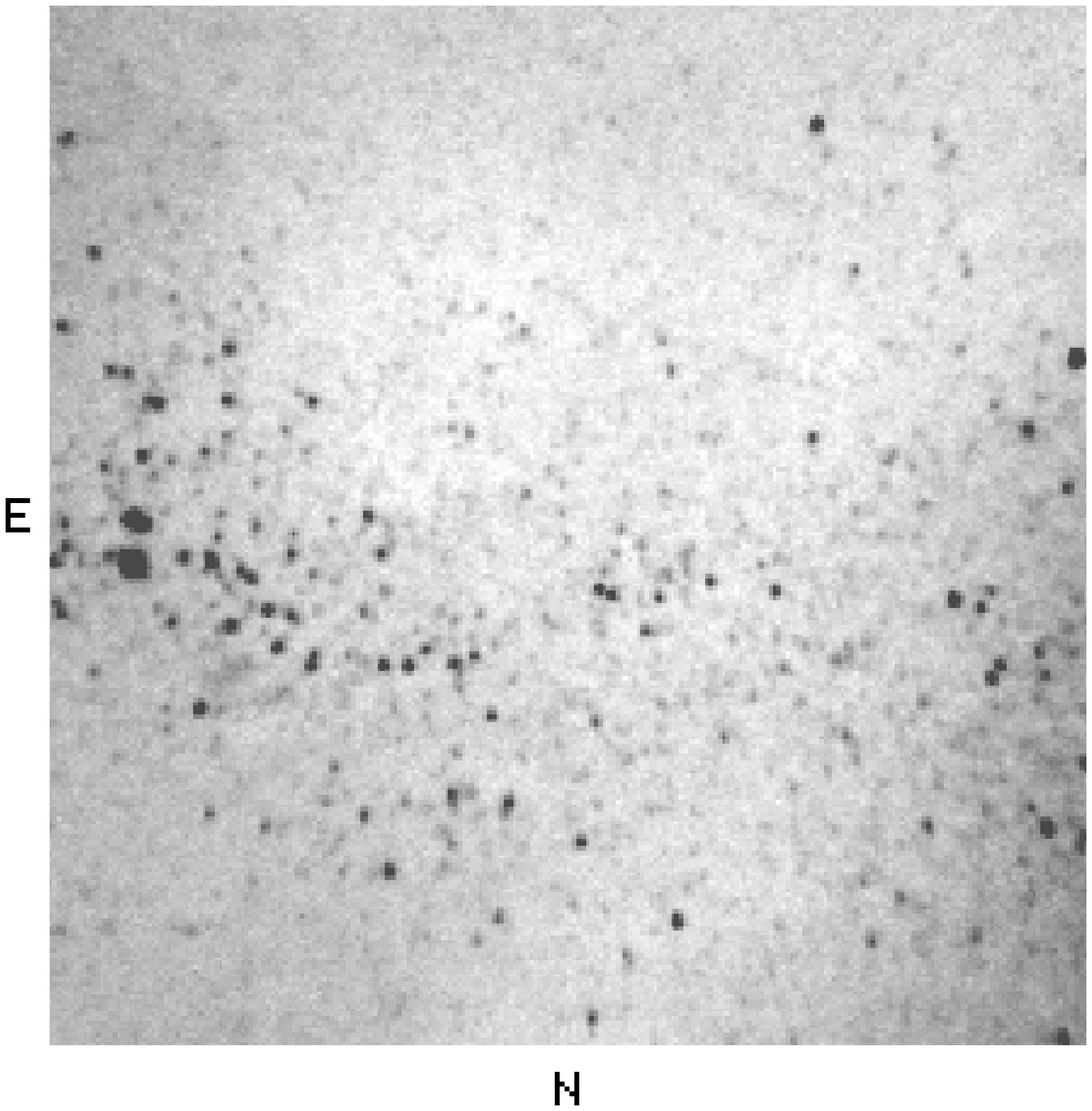}
\caption{Final $K$ image of NGC55 Field 2.}
\end{figure}

\begin{figure}
\plotone{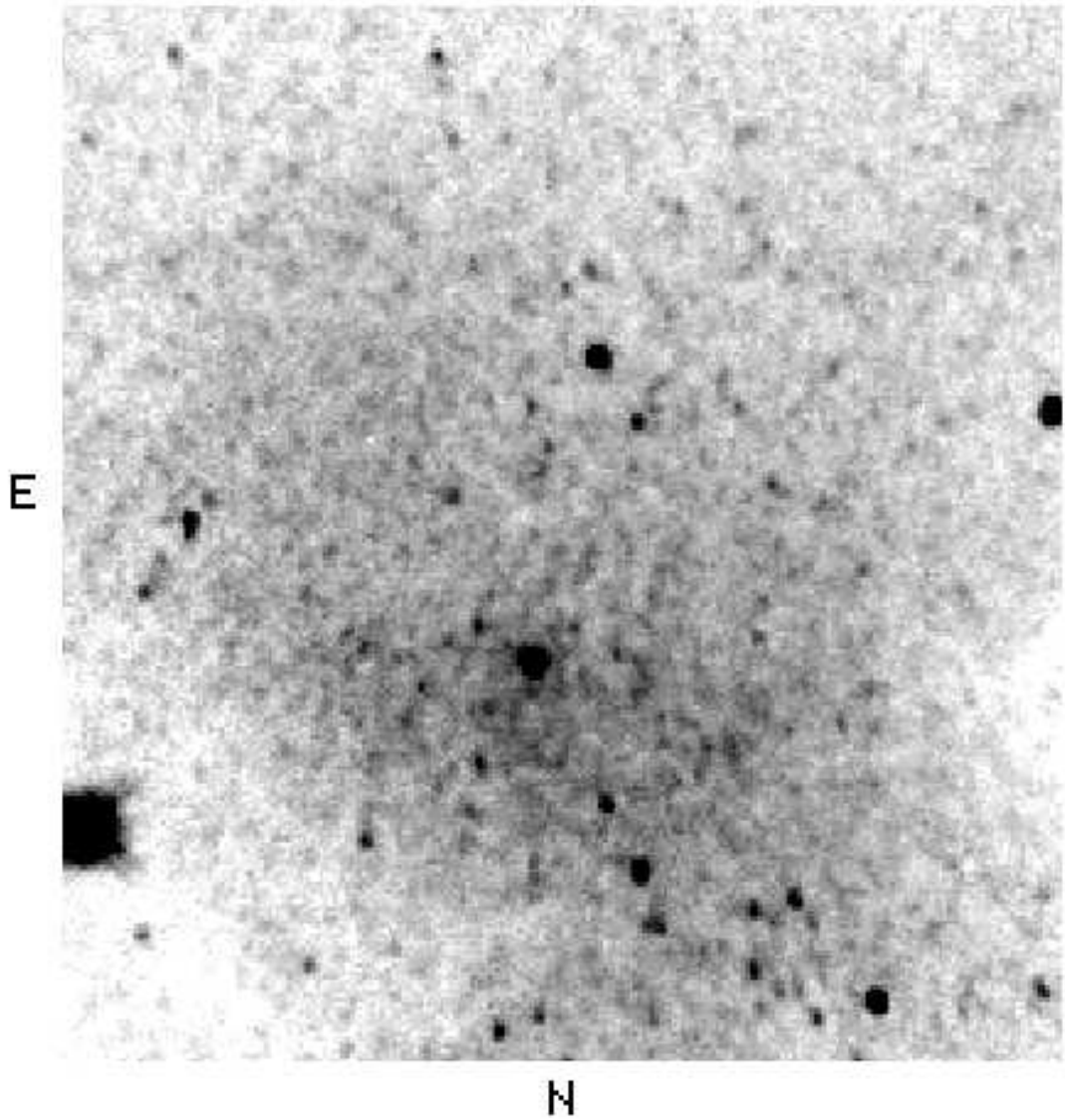}
\caption{Final $K$ image of NGC300 Field 1. The bulge is the bright 
stellar object slightly below and to the left of the field center.}
\end{figure}

\begin{figure}
\plotone{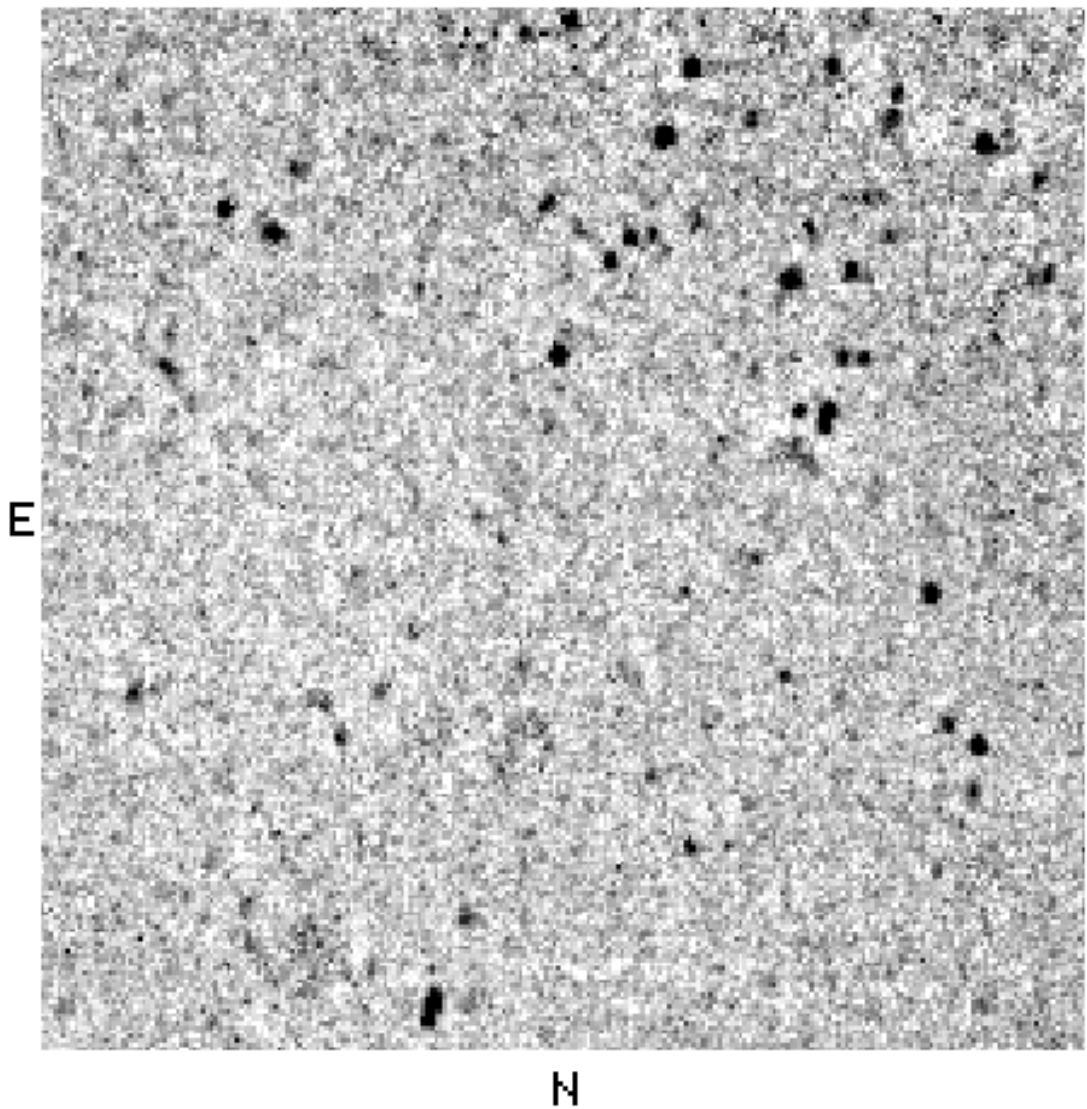}
\caption{Final $K$ image of NGC300 Field 2.}
\end{figure}

\begin{figure}
\plotone{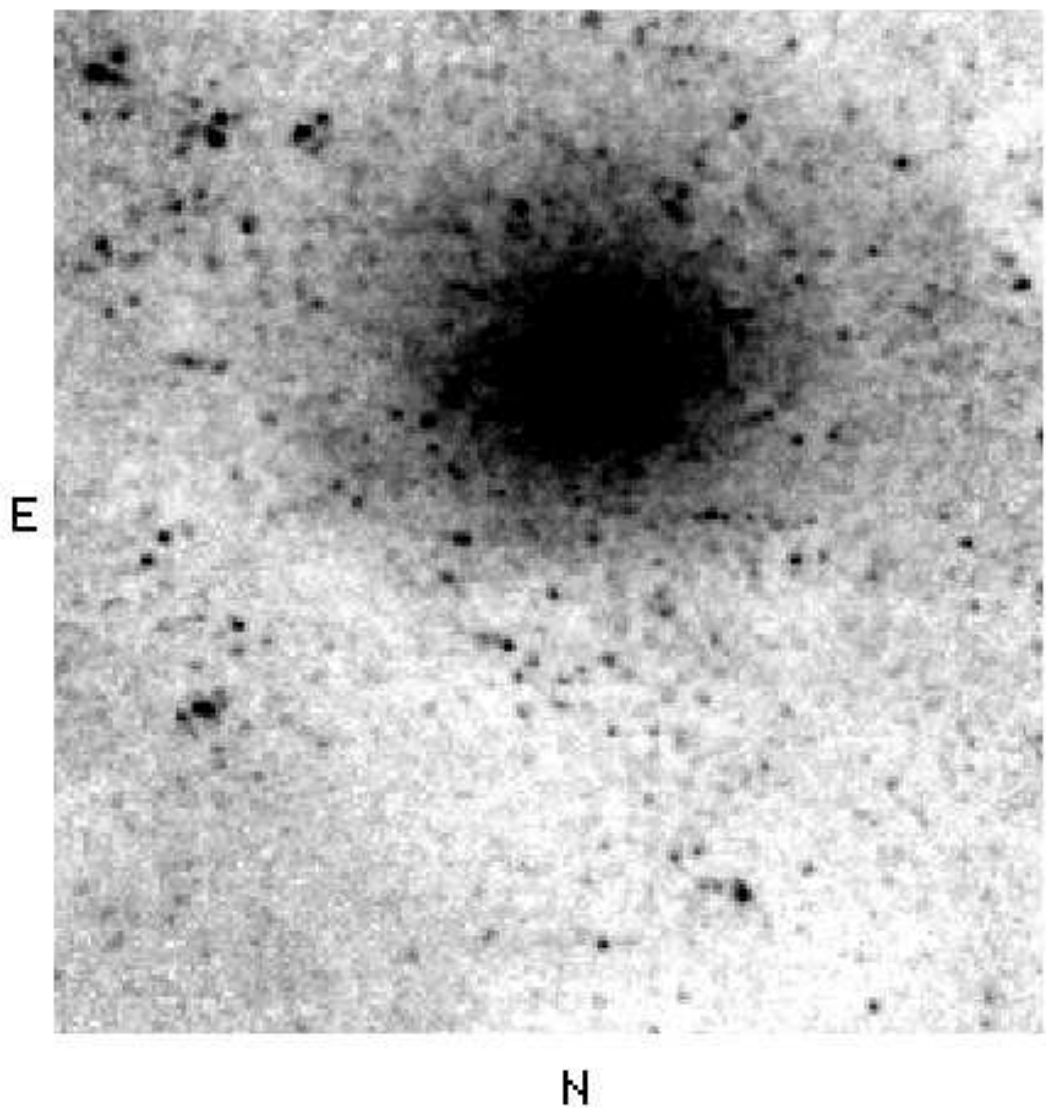}
\caption{Final $K$ image of NGC7793 Field 1.}
\end{figure}

\begin{figure}
\plotone{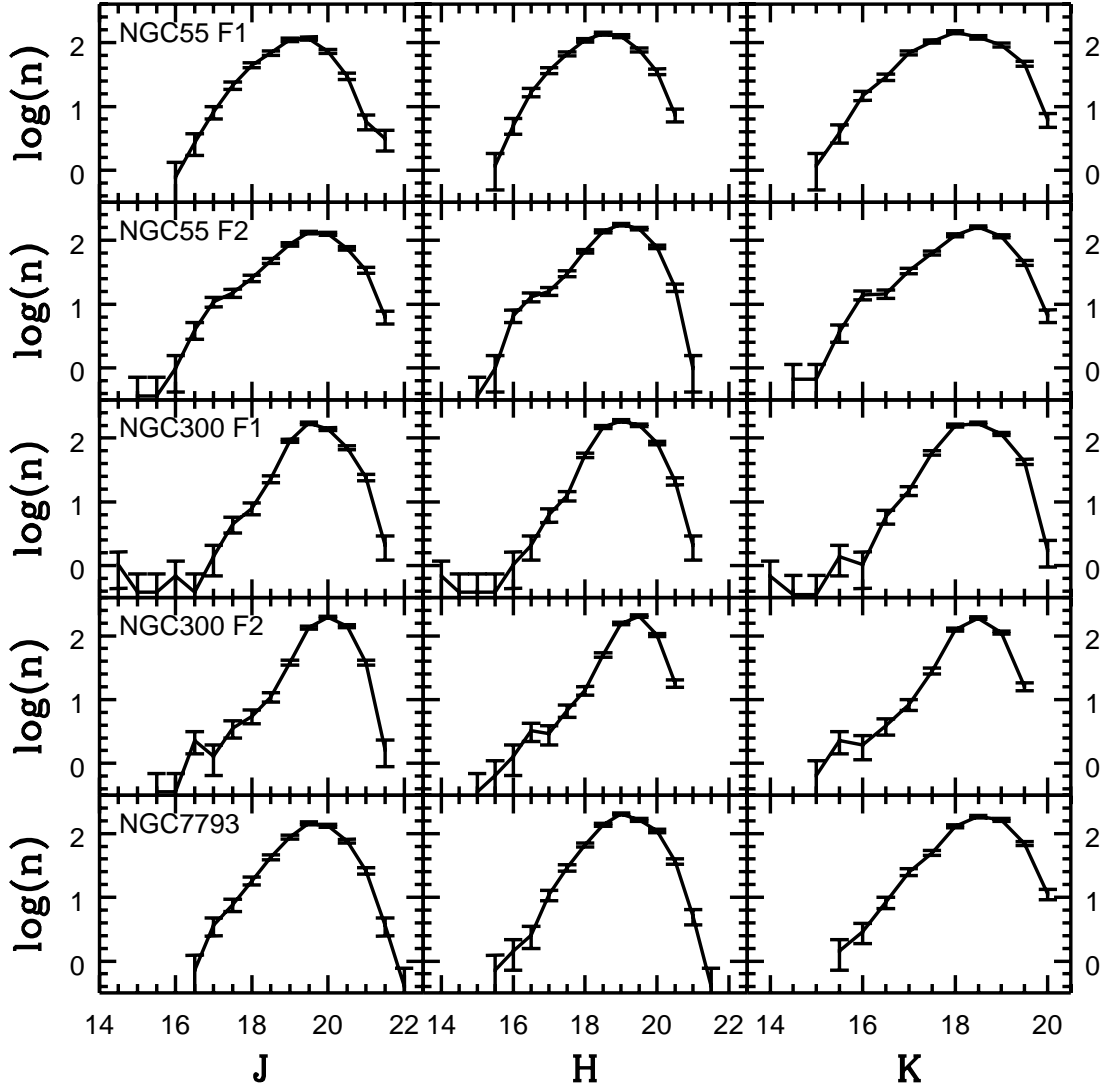}
\caption{$J, H,$ and $K$ LFs. $n$ is the number of stars per square arcmin per 
magnitude. The errorbars show the uncertainties introduced by 
counting statistics. Note that the $J, H,$ and $K$ LFs 
for a given field show many similarities. Moreover, while 
there is good field-to-field agreement within a given galaxy, 
significant galaxy-to-galaxy variations are clearly apparent.} 
\end{figure}

\begin{figure}
\plotone{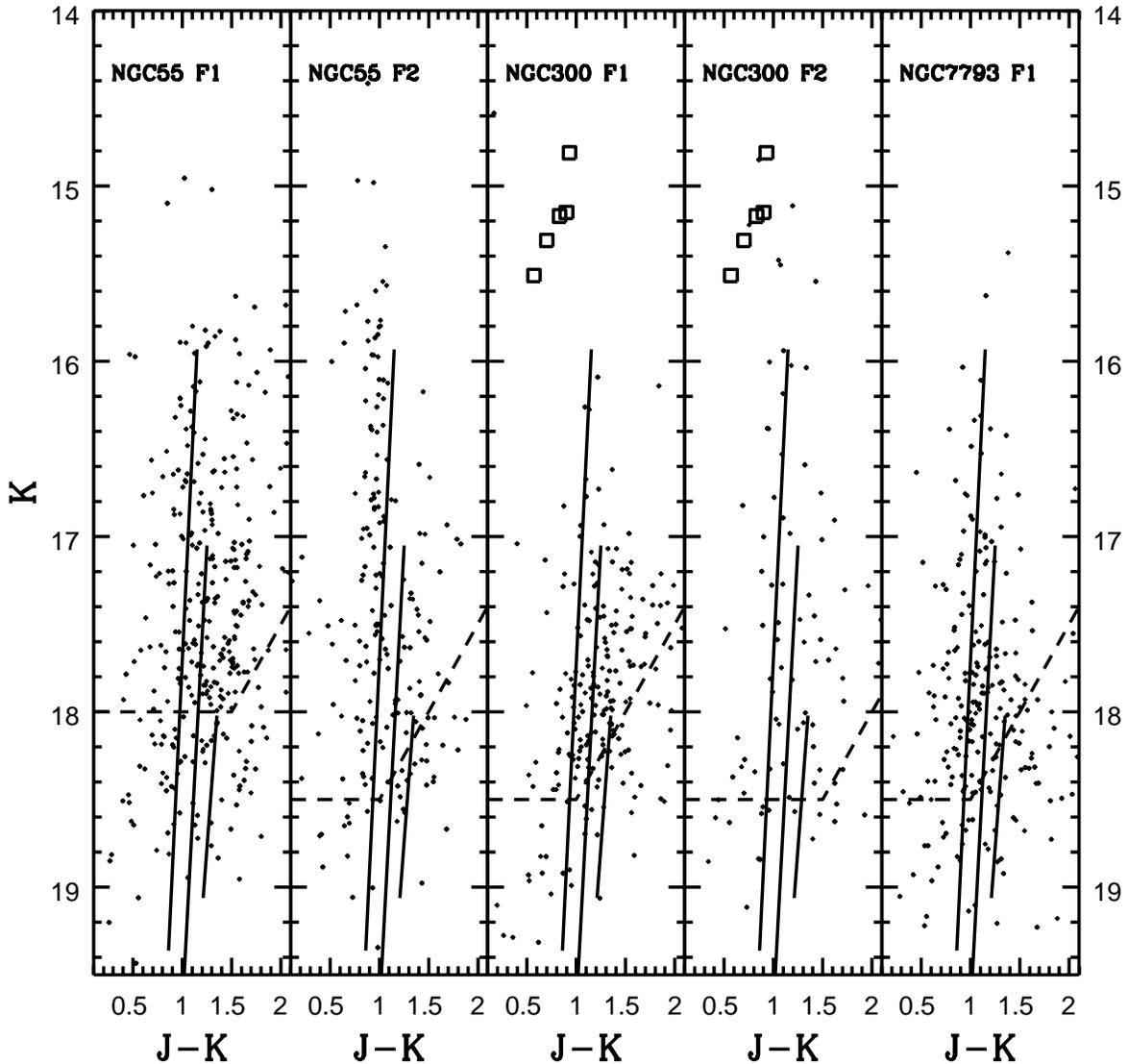}
\caption{$(K, J-K)$ CMDs. The open squares show photometric measurements of 
RSGs in NGC300 obtained by Humphreys \& Graham (1986). The dashed lines define 
the 100\% completeness limits. The solid lines are 
solar metallicity AGB sequences, formed by connecting the RGB-tip and 
AGB-tip points listed by Bertelli et al. (1994), for 
ages 0.1, 1.0, and 10 Gyr. A distance modulus of 26.0 has been assumed.}
\end{figure}

\begin{figure}
\plotone{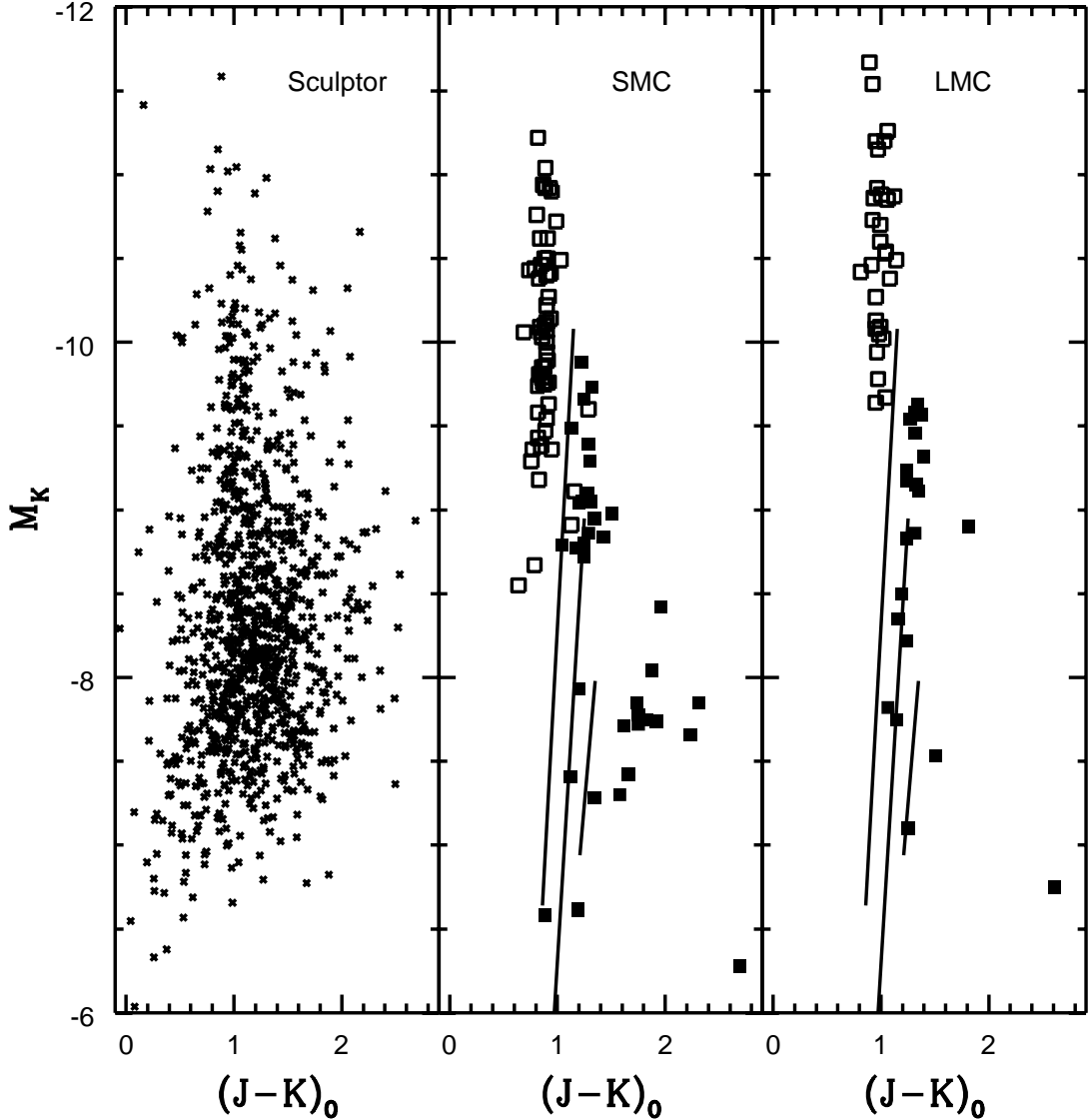}
\caption{The composite $(M_K, (J-K)_0)$ CMD of all five Sculptor fields, 
compared with CMDs of RSGs and LPVs in the SMC and LMC. 
The RSG measurements, shown as open squares, are from Tables 3 and 4 
of Elias et al. (1985), and only those stars with supergiant spectral-types 
listed in the third column of these Tables have been plotted. The 
AGB stars, shown as filled squares, are the LPVs listed 
in Tables 1 and 2 of Wood et al. (1983) that are confirmed AGB objects. 
In many cases Elias et al. (1985) and Wood et al. 
(1983) give measurements covering several epochs, and 
the points plotted are the mean of all values for a given object. The distance 
moduli of the LMC and SMC were assumed to be $\mu_0 = 18.5$ and 18.8, 
respectively. The Magellanic Cloud easurements were corrected 
for foreground reddening using the A$_B$ values given by Burstein \& Heiles 
(1984) and the reddening curve of Rieke \& Lebofsky (1985). The solid lines are 
solar metallicity AGB isochrones with ages 0.1, 1.0, and 10 Gyr, formed by 
connecting the RGB-tip and AGB-tip points listed by Bertelli et al. (1994).}
\end{figure}

\begin{figure}
\plotone{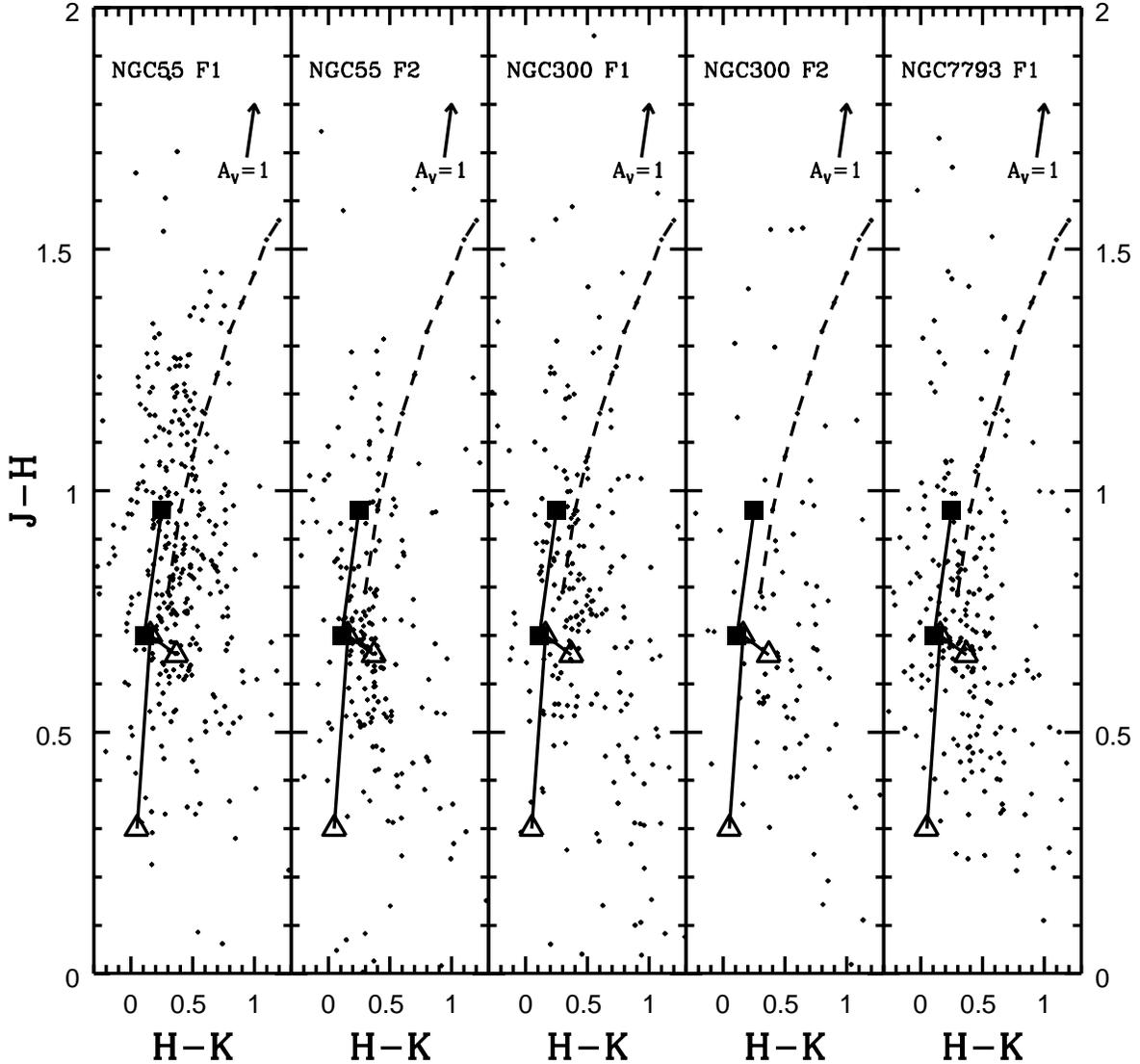}
\caption{$(J-H, H-K)$ two-color diagrams. The reddening vector predicted 
by the reddening law of Rieke \& Lebofsky (1985), and with a length 
corresponding to A$_V = 1$, is shown in the upper right 
hand corner of each panel. The line connecting the open 
triangles is the solar neighborhood dwarf relation listed in Table II 
of Bessell \& Brett (1988). The sequence shown here covers spectral 
types G0 to M6, and indicates that contamination from 
foreground stars is not an issue when $J-H \geq 0.7$. 
The line connecting the filled squares is the locus of 
RSGs in the Magellanic Clouds, based on data given by Elias et al. (1985). 
Finally, the dashed line shows the locus of LMC LPVs, based on data 
published by Wood et al. (1983, 1985). It is apparent that 
objects with $J-H \geq 0.95$ are likely AGB stars.}
\end{figure}

\begin{figure}
\plotone{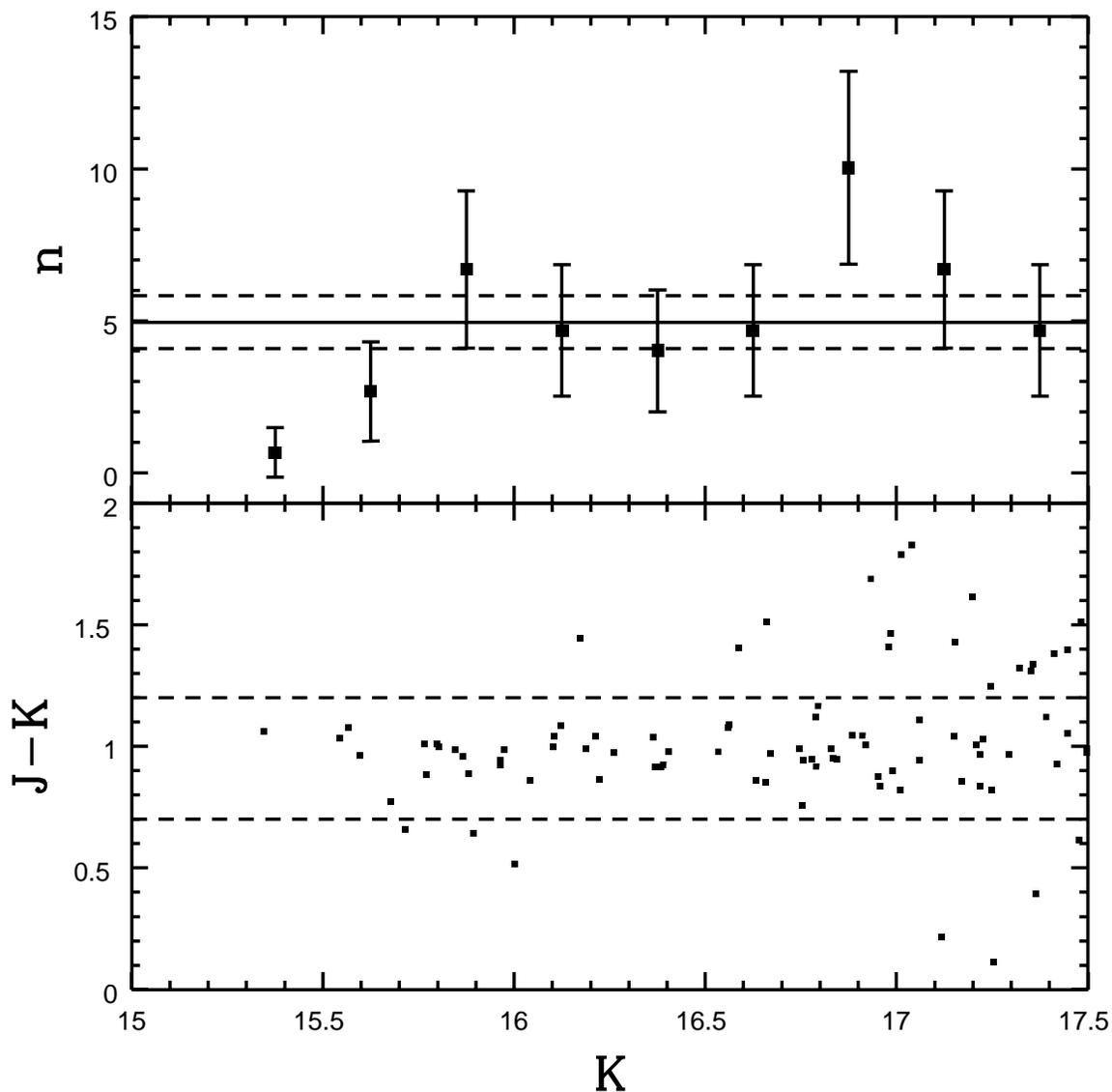}
\caption{The top panel shows the LF of stars in NGC55 Field 2 with $K \leq 
17.5$ and $0.7 \leq J-K \leq 1.2$. $n$ is the number of stars per square arcmin 
per magnitude interval. The error bars are uncertainties computed 
from Poisson statistics. The solid line is the mean of all the data points, 
and the dashed lines define the error in the mean. The lower panel shows 
the corresponding portion of the NGC55 Field 2 CMD, with the dashed lines 
defining the color boundaries used for computing the LF.} 
\end{figure}

\begin{figure}
\plotone{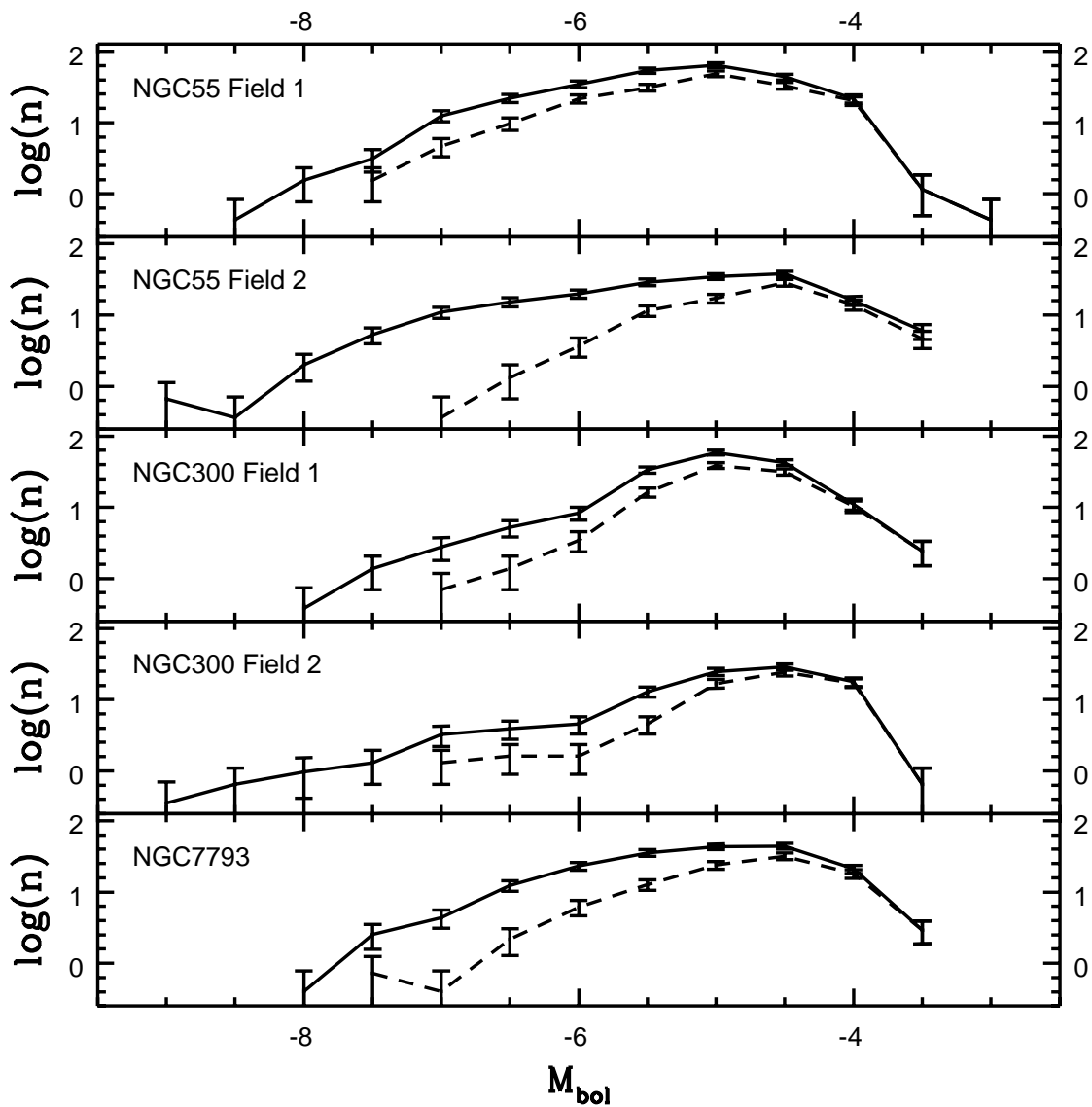}
\caption{The bolometric LFs of the five fields. $n$ is the number 
of stars per square arcmin per magnitude interval. The solid lines show 
the LFs for all stars, while the dashed lines show the LFs for objects 
with $(J-K) \geq 1.1$, which are LPVs. The error bars show 
uncertainties based on Poisson statistics.}
\end{figure}

\begin{figure}
\plotone{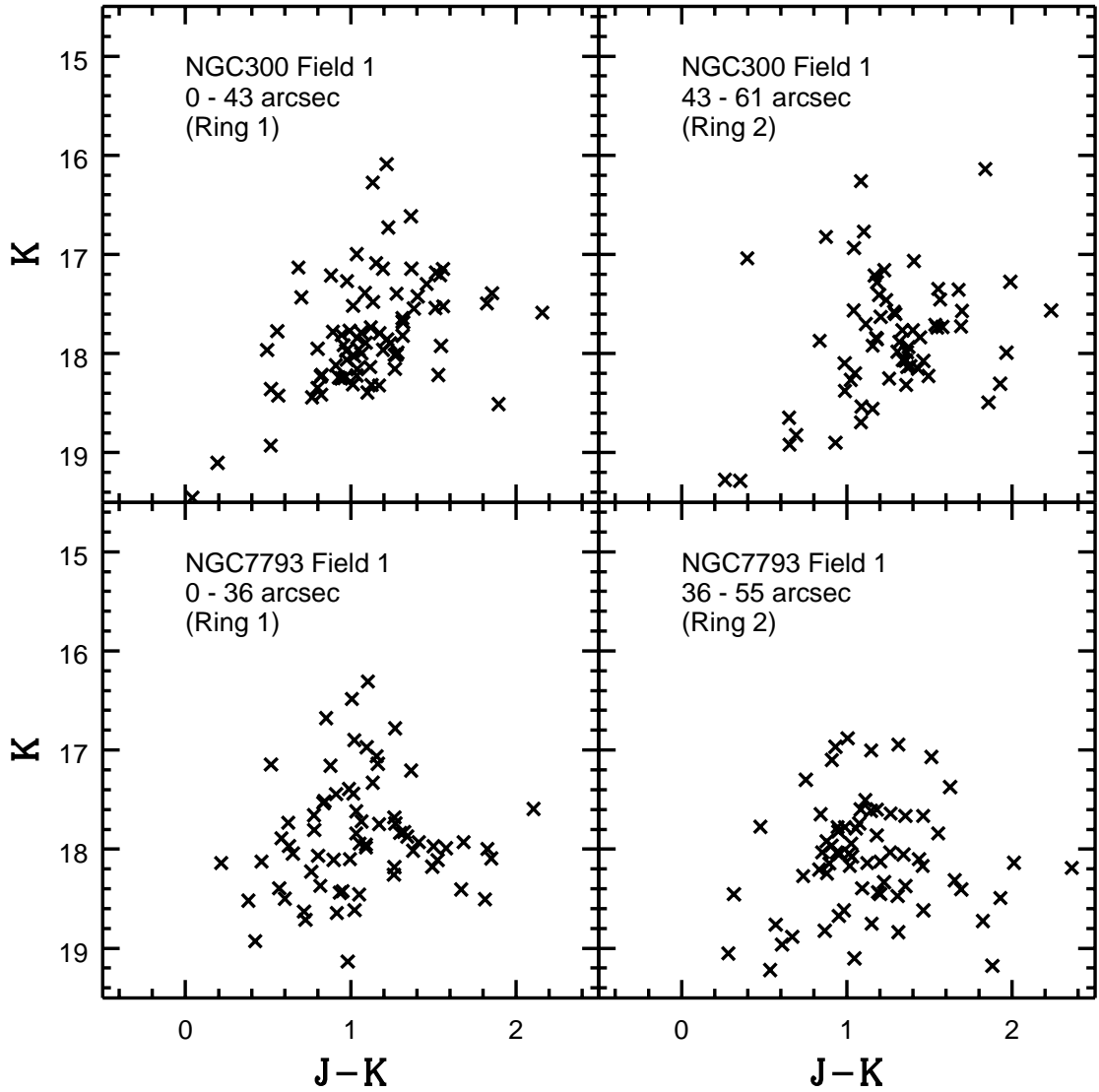}
\caption{CMDs for Rings 1 and 2 in Field 1 of NGC300 (top row) and NGC7793 
(bottom row). The radial intervals specified are measured from the galaxy 
centers.}
\end{figure}

\begin{figure}
\plotone{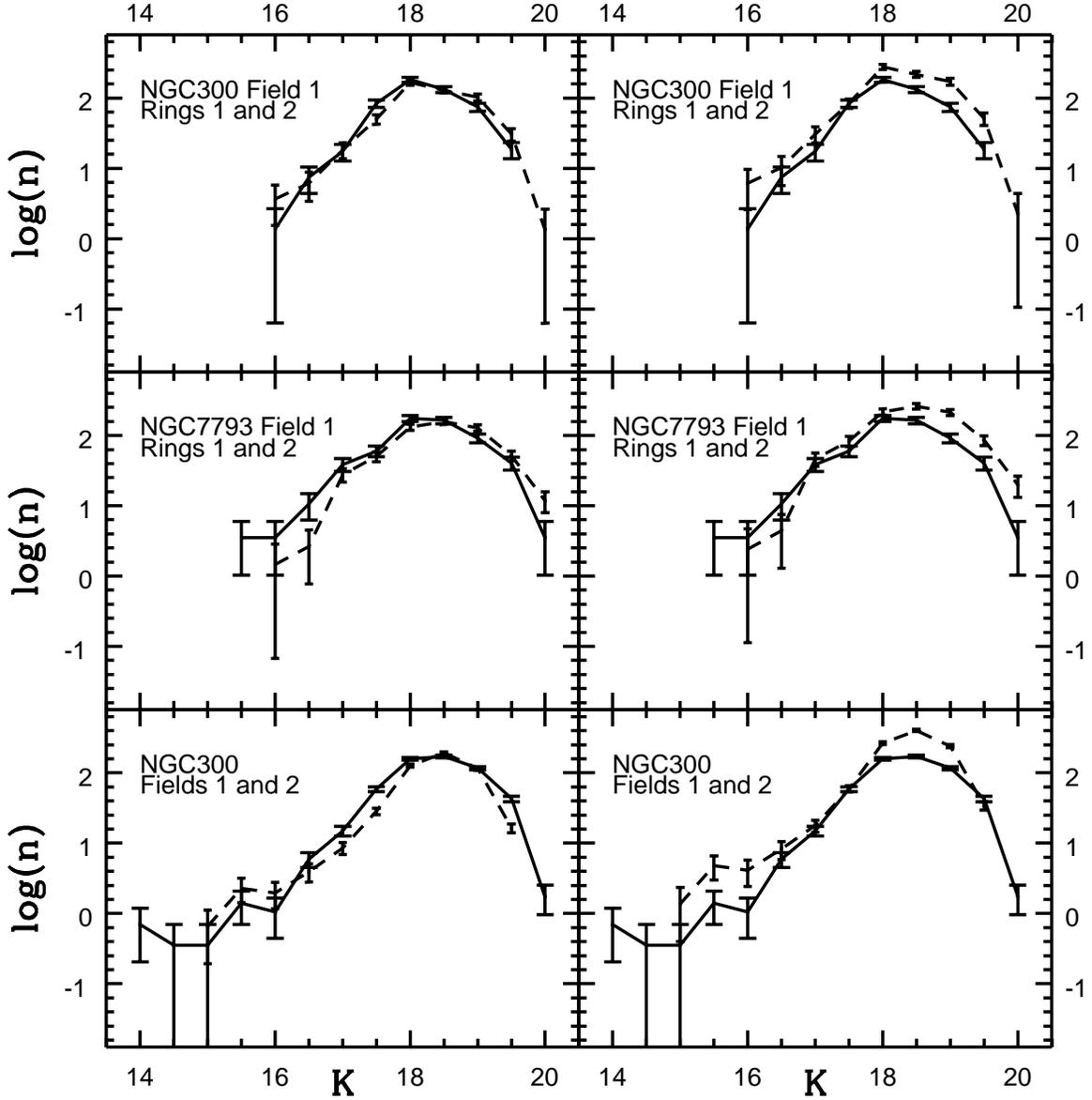}
\caption{$K$ LFs measured in Rings 1 and 2 of NGC300 Field 1 (top row) and 
NGC7793 Field 1 (middle row). The Ring 1 LFs are plotted with a solid line, 
while the Ring 2 LFs are shown as a dashed line. The bottom row compares the 
$K$ LFs of NGC300 Field 1 (solid line) and 2 (dashed line). $n$ is the number 
of stars per square arcmin per magnitude interval. The Ring 2 LFs in the right 
hand column of the top two panels have been shifted to match the Ring 1 mean 
surface brightnesses, based on the light profiles published by Carignan (1985). 
The Field 2 LF in the right hand column of the bottom panel has been shifted to 
match the Ring 2 mean surface brightnesses. The error bars show the 
uncertainties introduced by counting statistics.}
\end{figure}
\end{document}